\documentclass[fleqn,10pt]{wlscirep}
\usepackage[utf8]{inputenc}
\usepackage[T1]{fontenc}
\usepackage{lineno}
\usepackage[autostyle=true]{csquotes}
\usepackage{babel}
\usepackage{mathtools, amsmath}
\usepackage{bbold}
\usepackage{graphicx}
\usepackage{physics, siunitx}
\usepackage{booktabs, multirow}
\usepackage{caption}
\usepackage{subcaption}
\usepackage{breqn}
\usepackage[ruled,vlined]{algorithm2e}

\usepackage{blindtext}

\title{Quantum Compiling by Deep Reinforcement Learning}

\author[1,3]{Lorenzo Moro}
\author[2]{Matteo G. A. Paris}
\author[1]{Marcello Restelli}
\author[3,*]{Enrico Prati}

\affil[1]{Dipartimento di Elettronica, Informazione e Bioingegneria, Politecnico di Milano, Via Colombo 81, I-20133 Milano, Italy}
\affil[2]{Quantum Technology Lab, Dipartimento di Fisica Aldo Pontremoli, Università degli Studi di Milano, I-20133 Milano, Italy}
\affil[3]{Istituto di Fotonica e Nanotecnologie, Consiglio Nazionale delle Ricerche, Piazza Leonardo da Vinci 32, I-20133 Milano, Italy}
\affil[*]{corresponding author(s): Enrico Prati (enrico.prati@cnr.it)}

\begin{abstract}
The architecture of circuital quantum computers requires computing layers devoted to compiling high-level quantum algorithms into lower-level circuits of quantum gates. The general problem of quantum compiling is to approximate any unitary transformation that describes the quantum computation, as a sequence of elements selected from a finite base of universal quantum gates. The existence of an approximating sequence of one qubit quantum gates is guaranteed by the Solovay-Kitaev theorem, which implies sub-optimal algorithms to establish it explicitly. Since a unitary transformation may require significantly different gate sequences, depending on the base considered, such a problem is of great complexity and does not admit an efficient approximating algorithm. Therefore, traditional approaches are time-consuming tasks, unsuitable to be employed during quantum computation. We exploit the deep reinforcement learning method as an alternative strategy, which has a significantly different trade-off between search time and exploitation time. Deep reinforcement learning allows creating single-qubit operations in real time, after an arbitrary long training period during which a strategy for creating sequences to approximate unitary operators is built. The deep reinforcement learning based compiling method allows for fast computation times, which could in principle be exploited for real-time quantum compiling.
\end{abstract}
\begin{document}

\flushbottom
\maketitle

\thispagestyle{empty}

Quantum computation takes place at its lowest level by means of physical operations described by unitary matrices acting on the state of qubits. However, gate-model quantum computers may in practice provide just a limited set of transformations according to the constraints in their architecture~\cite{linke:experimentalComparison, maslov:ionCompilationTech,leibfried:ionOptimization,debnath:demonstrationIon}. Therefore, the computation is achieved as circuits of quantum gates, which are ordered sequences of unitary operators, acting on a few qubits at once. Although the Solovay Kitaev theorem~\cite{kitaev:theorem} ensures that any computations can be approximated, within an arbitrary tolerance, as a circuit based on a finite set of operators, there is no optimal strategy to establish how to compute such a sequence. The problem is known as quantum compiling and the algorithms to compute suitable approximating circuits as quantum compilers.

Every quantum compiler has its own trade-off between the length of the sequences, which should be as short as possible, the precompilation time, i.e., the time taken by the algorithm to be ready for use, and finally the execution time, i.e, the time the algorithm takes to return the sequence~\cite{zhiyenbayev:diffusive}. 

Previous works~\cite{barenco1995elementary, chuang:efficient, kitaev2002classical, zhiyenbayev:diffusive}, mostly based on the Solovay-Kiteav theorem, addressed the problem by providing algorithms that return the approximating sequence with lengths and execution-times that scale polylogarithmic as $\mathcal{O}(\log^c(1/\delta))$, where $\delta$ is the accuracy and $c$ is a constant between $3$ and $4$. For instance, the Dawson-Nielsen (\textsc{DNSK}) formulation~\cite{dawson:solovay} provides sequences of length $\mathcal{O}(\log^{3.97}(1/\delta))$ in a time of $\mathcal{O}(\log^{2.71}(1/\delta))$. Additional performance gains can be achieved by selecting unique sets of quantum gates~\cite{zhiyenbayev:diffusive}, reaching lengths that scale as $\mathcal{O}(\log^{\log(3)/\log(2)}(1/\delta))$ at the cost of increasing the pre-compilation time. 
Hybrid approaches involving a planning algorithm~\cite{davis2020towards}, in some cases boosted by deep neural networks~\cite{PhysRevLett.125.170501}, could achieve better performance. However, the planning algorithm raises the execution time, which could scales sub-optimally for high accuracy.
Despite the strategy considered, no algorithm can return sequence using less than $\mathcal{O}(\log(1/\delta))$ gates, as shown in~\cite{chuang:efficient} by a geometrical proof. 

While existing quantum compilers are characterized by high execution and precompilation times~\cite{zhiyenbayev:diffusive, dawson:solovay}, which make them impractical to compute during online operations, deep learning suggests an alternative approach. By exploiting a deep reinforcement learning algorithm, it would be in principle possible to train an agent to generalize how to map any unitary operator into a sequence of elementary gates within an arbitrary precision. A quantum compiler based on a deep neural network is characterized by a long precompilation time, performed only once, and no guarantee of always finding a suitable solution. However, writing the map into the deep neural network would enable to call online quantum compilers, greatly reducing execution time and allowing dynamical programming of a gate-model quantum computer. This method is irrespective of the specific hardware constraining the physical operations on the qubits and it can be easily scaled to exploit precompiled basis of gates.


Recently, deep learning has been successfully applied to physics~\cite{fosel2018reinforcement, dunjko2018machine, Sarma2019MLQF, carleo2019machine}, where unprecedented advancements have been achieved by combining reinforcement learning~\cite{sutton1998reinforcement} with deep neural networks into deep reinforcement learning (\textsc{DRL}). \textsc{DRL}, thanks to its ability to identify strategies for achieving a goal in complex configuration spaces without prior knowledge of the system~\cite{mnih2015human,Melnikov2017,Nautrup2018,Sweke2018,Reddy2016,Tor2017}, has recently been proposed for the control of quantum systems~\cite{fosel2018reinforcement,August2018,Niu2018,Retamal2018,niu2019universal,andreasson2019quantum,prati2017quantum}. 
In this context, some of us previously applied deep reinforcement learning to control and initialize qubits by continuous pulse sequences ~\cite{porotti:CTAP,porotti2019reinforcement} for coherent transport by adiabatic passage (\textsc{CTAP})~\cite{ferraro2015coherent} and by digital pulse sequences for stimulated Raman passage (\textsc{STIRAP})~\cite{paparelle2020digitally}, respectively. Furthermore, it has proven effective as a control framework for optimizing the speed and fidelity of quantum computation~\cite{google:universalcontrol} and in control of quantum gates~\cite{RL:gatecontrol}. 


In this work, we propose a novel approach to quantum compiling, exploiting deep reinforcement learning to approximate, with competitive tolerance, single-qubit unitary operators as circuits made by an arbitrary initial set of elementary quantum gates.
As prominent examples, we show how to steer quantum compiling for small rotations of $\pi/128$ around the three-axis of the Bloch sphere and for the Harrow-Recht-Chuang efficiently universal gates~\textsc{(HRC)}~\cite{chuang:efficient}, by employing two alternative \textsc{DRL} algorithms depending on the nature of the base. After training, agents can generate single-qubit logic circuits within a competitive tolerance of $0.99$ average-gate fidelity (\textsc{AGF}).
The strategy we used consists of generating a uniform distribution of single-qubit unitary matrices, where to sample the training targets. The agents are not told how to approximate such targets, but instead, they are asked to establish a suitable policy to complete the task. The agents' final performance is then measured using a validation set of unitary operators not previously seen by the agent.

The \textsc{DRL} agents prove to capture the intricate structure of the unitary matrix space, discovering approximating circuits of unitary single-qubit operations and providing a viable approach to perform real-time quantum compiling, once the neural network is trained.

\section*{Results}

\subsection*{Deep reinforcement learning}
Deep reinforcement learning is a subset of machine learning that exploits deep neural networks to learn optimal policies in order to achieve specific goals in decision-making problems~\cite{tognetti2009batch, niu2019universal, castelletti2013multiobjective}. Such techniques can be remarkably effective in high-dimensional control tasks and to address problems where limited or no prior knowledge of the configuration space of the system is available.

The fundamental assumptions and concepts in the reinforcement learning theory are built upon the idea of continuous interactions between a decision-maker called agent and a controlled system named environment, typically defined in the form of a Markov decision process (\textsc{MDP})~\cite{barto:book}. According to a policy function that fully determines its behavior, the former interacts with the latter at discrete time-steps, performing an action based on an observation related to the current state of the environment. Therefore, the environment evolves changing its state and returning a reward signal, that can be interpreted as a measure of the adequateness of the action the agent has performed. The only purpose of the agent is to learn a policy to maximize the reward over time. The learning procedure can be a highly time-consuming task, but it has to be performed once. Then, it is possible to exploit the policy encoded in the deep neural network, with low computational resources in minimal time.

\subsection*{Deep reinforcement learning as quantum compiler}
The quantum compilation is a fundamental problem in the quantum computation theory, consisting of approximating any unitary transformation as a finite sequence of unitary operators $A_j$ chosen from a universal set of gates $\mathcal{B}$.

In this work, we ask the agent to approximate any single-qubit unitary matrix~$\mathcal{U}$, within a fixed tolerance $\varepsilon$. Therefore, the goal of the agent is to find a unitary matrix $U_n=\prod_{j=1}^{n}{A_j}$, resulting from the composition of the elements in the sequence, that is sufficiently close to $\mathcal{U}$. Although the \textsc{DRL} framework allows exploiting any distance between matrices to evaluate the accuracy of the solutions, the average gate fidelity is wildly used for the purpose, mainly due to the modest computational demands needed to compute it. Alternative choices are possible, such as the diamond norm~\cite{kitaev:diamondNorm, watrous:diamond}.

\begin{figure}
    \includegraphics[width=\textwidth]{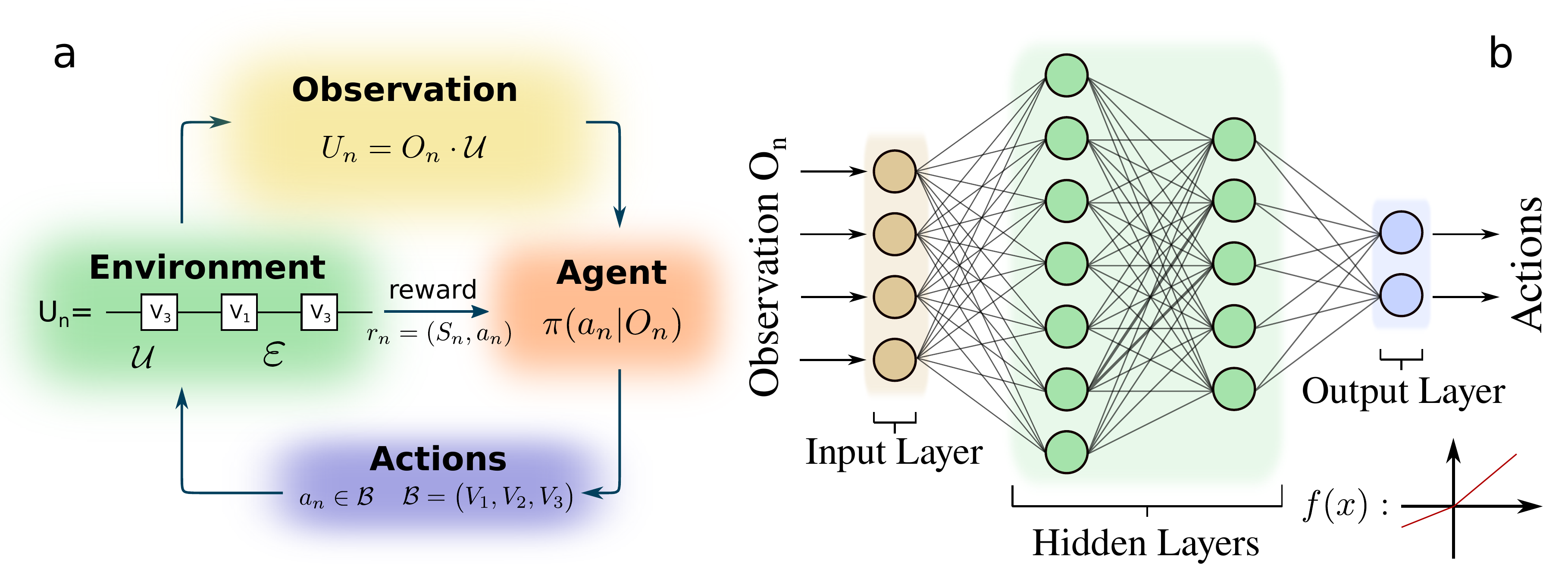}
    \caption{\textbf{The deep reinforcement learning architecture}. \textbf{a} The \textsc{DRL} environment can be described as a quantum circuit modeled by the approximating sequence $U_n$, the fixed tolerance $\varepsilon$ and the unitary target to approximate $\mathcal{U}$, that generally changes at each episode. At each time-step $n$ the agent receives the current observation $O_n$ and based on that information it chooses the next gate to apply on the quantum circuit. Therefore, the environment returns the real-valued reward $r_n$ to the agent. \textbf{b}. The policy $\pi$ of the agent is encoded in a deep neural network. At each time-step, the \textsc{DNN} receives as input a vector made by the real and imaginary part of the observation $O_n$. Such information is processed by the hidden layers and returned through the output layer. The neurons in the output layer are associated with the action the agent will perform in the next time-step. In the bottom-right corner is reported an example of the non-linear activation function $\textsc{RELU}$.}
    \label{fig:fig0}
\end{figure}

In the framework of quantum compiling, the environment consists of a quantum circuit that starts as the identity at the beginning of each episode. It is built incrementally at each time-step by the agent, choosing a gate from $\mathcal{B}$ according to the policy $\pi$ encoded in the deep neural network, as shown in Figure~\ref{fig:fig0}. Therefore, the available actions that the agent can perform correspond to the gates in the base~$\mathcal{B}$.

The observation used as input at time-step $n$ corresponds to the vector of the real and imaginary parts of the elements of the matrix $O_n$, where $\mathcal{U} = U_n \cdot O_n$. Such representation encodes all the information needed by the agent to build a suitable approximating sequence of gates, i.e. the current composition of gates and the unitary target to approximate. No information on the tolerance is given to the agent since it's fixed and thus it can be learned indirectly during the training.

Designing a suitable reward function is remarkably challenging, potentially leading to unexpected or unwanted behavior if not defined accurately. Therefore, two reward functions have been designed depending on the different characteristics of the gate base considered, which can be identified as quasi-continuous-like sets of small rotations and discrete sets. The former are inspired by gates available on superconductive and trapped ions architecture~\cite{linke:experimentalComparison, debnath:demonstrationIon, maslov:ionCompilationTech}, where the latter are standard set of logic gates, typically used to write quantum algorithms, e.g. the Clifford+T library~\cite{Chuang:book, tolar:cliffordReasons}. Both reward functions are negative at each time-step, so that the agent will prefer shorter episodes. More details on the rewards are given in the Supplementary Note~$2$.

In this work we exploit Deep Q-Learning (\textsc{DQL})~\cite{watkins:qLearning} and Proximal Policy Optimization (\textsc{PPO})~\cite{schulman:PPO} algorithms to train the agents, depending on the reward function. Such algorithms differs in many aspects, as described in the Supplementary Notes $1.1$ and $1.2$ respectively. The former is mandatory for the case of sparse reward, since such reward requires an off-policy methods to be exploited, while the latter has been chosen for its robustness and tunability.

\subsection*{Training neural networks for approximating a single-qubit gate}
\label{sub:fixedTarget}
To demonstrate the exploitation of \textsc{DRL} as a quantum compiler, we first considered the problem of decomposing a single-qubit gate $\mathcal{U}$, into a circuit of unitary transformations that can be implemented directly on quantum hardware. The base of gates corresponds to six small rotations of $\pi/128$ around the three-axis of the Bloch sphere, i.e $\mathcal{B}=\bigl(R_{\hat{x}}(\pm\frac{\pi}{128}), \, R_{\hat{y}}(\pm\frac{\pi}{128}), \, R_{\hat{z}}(\pm\frac{\pi}{128})\bigr)$.

It is essential to choose the tolerance $\varepsilon$ and the fixed target accurately to appreciate the learning procedure. The former should be small enough and the latter sufficiently far from the identity not to be solved by chance. However, if the target is too difficult to approximate, the agent will fail and no learning occurs. To be sure that at least one solution does exist, we build $\mathcal{U}$ as a composition of $87$ elements selected from $\mathcal{B}$. The resulting unitary target is:
\begin{equation}
    \mathcal{U} =
    \begin{pmatrix*}[r]
    0.76749896-0.43959894 \rm{i} & -0.09607122+0.45658344 \rm{i} \\
    0.09607122+0.45658344 \rm{i} & 0.76749896+0.43959894 \rm{i}
    \end{pmatrix*}.
    \label{eq:fixedTargetMatrix}
\end{equation}

We tested a non-learning agent that acts randomly to ensure not to deal with a trivial task, setting the tolerance at $0.99$ average gate fidelity and limiting the maximum length of the episode at $130$. No solution was found after $10^4$ episodes. Then, the problem was addressed by exploiting a \textsc{DQL} agent, using the same thresholds for the tolerance and the length of the episode. We exploited the dense reward function: 
\begin{equation}
\label{eq:denseReward}
    r(S_n, a_n)=
        \begin{cases}
         (L-n)+1                            & \text{if $d(U_n, \mathcal{U})<\varepsilon$} \\
         -d(U_n, \mathcal{U})/L               & \text{otherwise}
        \end{cases}\\
\end{equation}
where $L$ is the maximum length of the episode, $a_n$, $S_n$ and $d(U_n, \mathcal{U}$) are the action performed, the state of the environment and the distance between the target and the approximating sequence at time $n$ respectively. Such reward performs adequately if small rotations are used as base only.

\begin{figure}
    \centering
    \includegraphics[width=\textwidth]{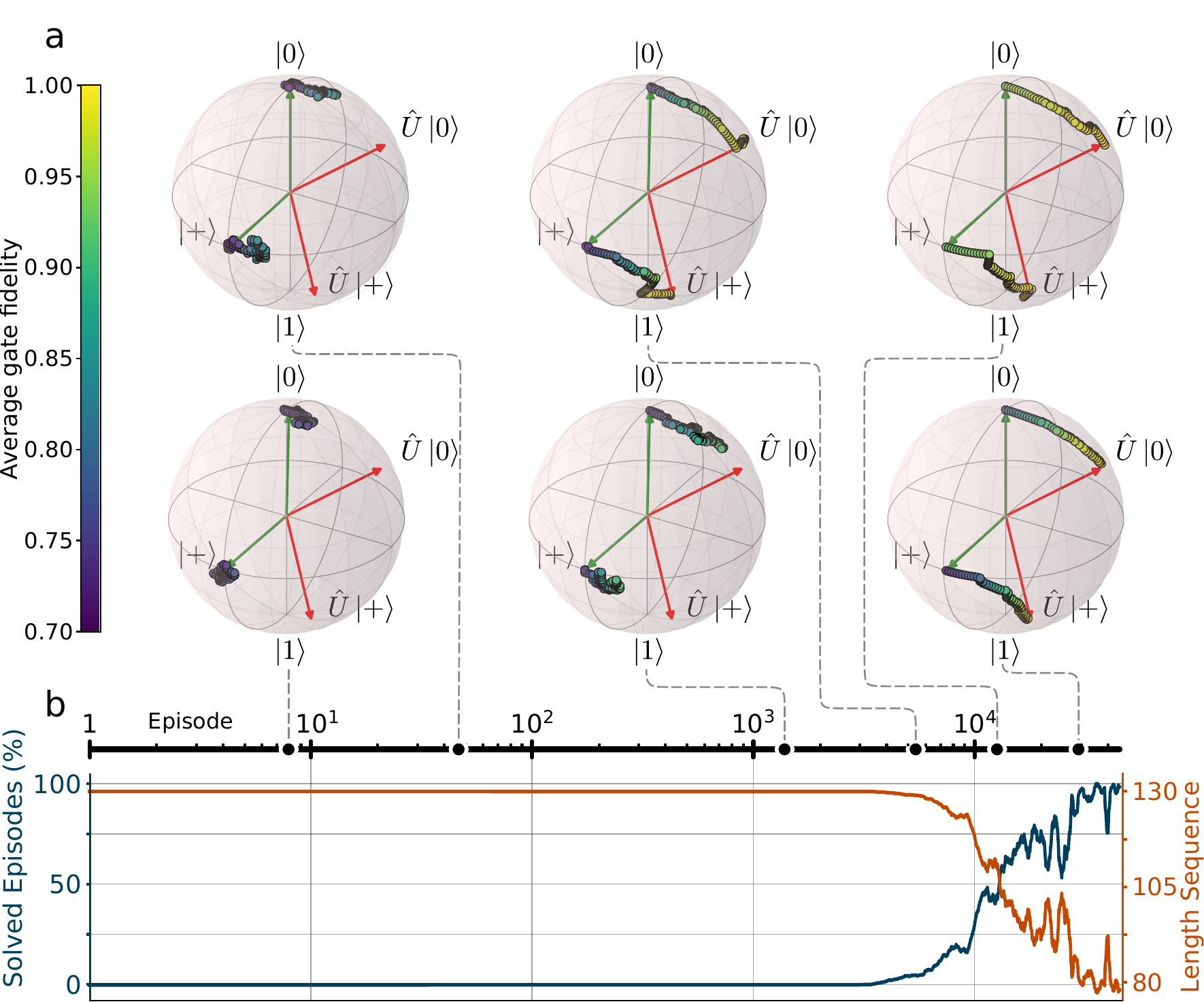}
    \caption{\textbf{The deep reinforcement learning agent learns how to approximate a single-qubit gate}. \textbf{a}, Best sequences of gates discovered by the agent during the training at different epochs. The timestamps indicate the time at which they were found for the first time. Each approximating sequence is represented by two trajectories of states (colored points) on the Bloch sphere. They are obtained by applying the unitary transformations associated with the circuit at the time-step $n$ on two representative states, namely $\ket{0}$ and $\ket{+}$ respectively. The agent is asked to transform the starting state (green arrows) in the corresponding ending state (red arrows), i.e. $\ket{0}$ to $\mathcal{U} \ket{0}$ and $\ket{+}$ to $\mathcal{U} \ket{+}$ respectively. \textbf{b} Performance of the agent during training. The plot represents the percentage of episodes for which the agent was able to find a solution (blue line) and the average number of the sequence of gates (orange line). The agent learns how to approximate the target after about $10^4$ episodes. Next, it continues to improve the solution over time.}
    \label{fig:fixed_target_learning}
\end{figure}

Table~\ref{tab:fixedTarget} reports some additional information about the networks architecture and the hyperparameters set, while Figure~\ref{fig:fixed_target_learning} shows the performance and the solutions found by the agent during the training time. The agent learns how to approximate the target after about $10^4$ episodes, while improving the solution over time. At the end of the learning, the agent discovered an approximating circuit made by $76$ gates only, within the target tolerance.

\subsection*{Quantum compiling by rotations operators}
The \textsc{DRL} approach can be generalized to the quantum compilation of a larger class of unitary transformations. Instead of limiting to approximating one matrix only, we aim at exploiting the knowledge of a trained agent to approximate any single-qubit unitary transformation, without requiring additional training. Therefore, we used as training targets Haar unitary matrices, since they form an unbiased and a general data set which is ideal to train neural networks, as described in the Methods section. If additional information on the type and distribution of targets is known, it is possible to choose a different set of gates for training, potentially increasing the performance of the agent as described in the Supplementary Note~$3.1$.

Such task is tougher to solve compared to the fixed-target problem. Therefore, we exploited the Proximal Policy Optimization algorithm (\textsc{PPO})~\cite{schulman:PPO} being more robust and easy to tune than \textsc{DQL}. We fixed the tolerance $\varepsilon$ at $0.99$ \textsc{AGF} and limited the maximum length of the approximating circuits (time-step per episode) at $300$ gates, as reported in Table~\ref{tab:rotationsHaar}. Figure~\ref{fig:composition_PPO_DQN}b shows the performance of the agent during the training time (blue lines). The agent starts to approximate unitaries after $10^5$ episodes, but it requires much more time to achieve satisfactory performance.  

We tested the performance of the agents at the end of the learning, using a validation set of $10^6$ Haar unitary targets. The agent is able to approximate more than $96\%$ of the targets within the tolerance requested. Complete results are reported in Table~\ref{tab:finalResults}.

\subsection*{Quantum compiling by the \textsc{HRC} efficiently universal base of gates}
In order to fully exploit the power of \textsc{DRL}, we now turn to the problem of compiling single-qubit unitary matrices using a base of discrete gates. The agent can perform the set of \textsc{HRC} efficiently universal base of gates~\cite{chuang:efficient}:
\begin{equation}
        V_1 = \frac{1}{\sqrt{5}}
    \begin{pmatrix}
        1 & 2\rm{i} \\
        2\rm{i} & 1
    \end{pmatrix} \qquad
    V_2 = \frac{1}{\sqrt{5}}
    \begin{pmatrix}
        1 & 2 \\
        -2 & 1
    \end{pmatrix} \qquad
    V_3 = \frac{1}{\sqrt{5}}
    \begin{pmatrix}
        1+2\rm{i} & 0 \\
        0 & 1-2\rm{i}
    \end{pmatrix}
\end{equation}
Such unitary matrices implement quantum transformations that are very different from the ones performed by small rotations. The agent has to learn how to navigate in the high-dimensional space of unitary matrices, exploiting counterintuitive movements that could lead it close to the target at the last time-step of the episode only. Therefore, the dense reward function~\ref{eq:denseReward} is no longer useful to guide the agent towards the targets. We exploited a "sparse" reward (binary reward):
\begin{equation}
    \label{eq:sparseReward}
        r(S_n, a_n)=
        \begin{cases}
         \mathrlap0\hphantom{-d(U_n, \mathcal{U})/L}      & \text{if $d(U_n, \mathcal{U})<\varepsilon$} \\
         -1/L   & \text{otherwise}.
        \end{cases}
\end{equation}
Such function lowers the reward of the agent equally for every action it takes, bringing no information to the agent on how to find the solution. Therefore, it requires advanced generalization techniques to be effective, such as Hindsight Experience Replay (\textsc{HER})~\cite{andrychowicz:HER}. Since \textsc{HER} requires an off-policy reinforcement learning algorithm, we chose the \textsc{DQL} agent to address the problem. 

\begin{figure}
    \centering
    \includegraphics[width=\textwidth]{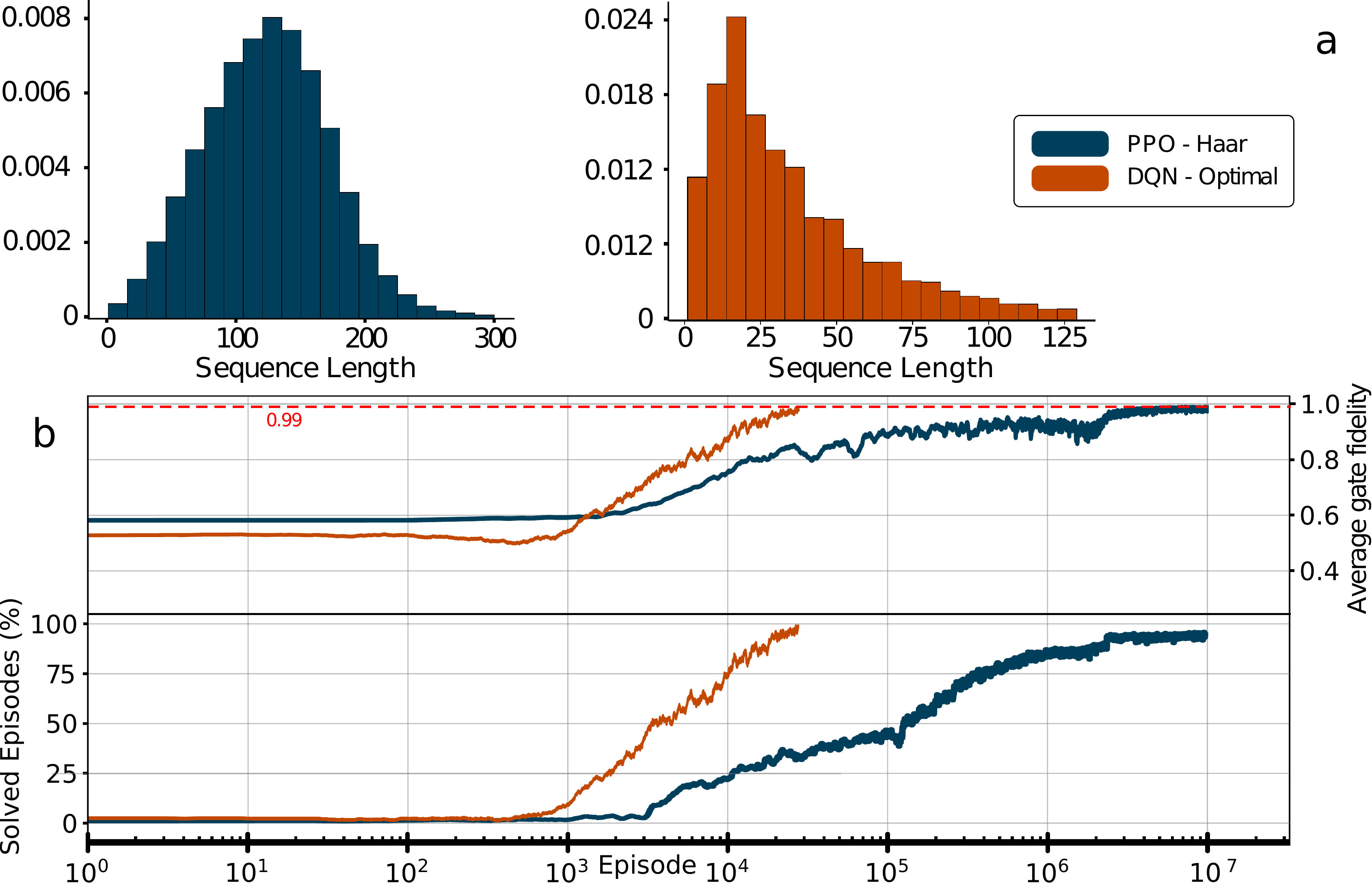}
    \caption{\textbf{Deep reinforcement learning agents learn how to approximate single-qubit unitaries using different base of gates}. A PPO agent (blue color) and a \textsc{DQL+HER} agent (orange color) were trained to approximate single-qubit unitaries using two different base of gates, i.e. six small rotations of $\pi/128$ around the three-axis of the Bloch sphere and the $\textsc{HRC}$ efficient base of gates, respectively. The tolerance was fixed to $0.99$ average gate fidelity. \textbf{a}, The length distributions of the gates sequences discovered by the agents at the end of the learning. The $\textsc{HRC}$ base generates shorter circuits as expected. \textbf{b}, Performance of the agent during training on the tasks.}
    \label{fig:composition_PPO_DQN}
\end{figure}

We fixed the tolerance $\varepsilon$ at $0.99$ \textsc{AGF} and limited the maximum length of the approximating circuits at $130$ gates. Figure~\ref{fig:composition_PPO_DQN} shows the performance of the agent during the training time (orange lines) and the length distribution of the solved sequences obtained using a validation set of $10^6$ Haar random unitaries. 
Although the agent receives a non-informative reward signal at each time-step, it surprisingly succeeds to solve roughly more than $95\%$ of the targets, using on average less than $36$ gates as reported in Table~\ref{tab:finalResults}. It is worth noting that the agent can build significantly shorter circuits, compared to the case of rotation matrices. It is not unexpected, since the \textsc{HRC} base allows to explore the space of unitary matrices quite efficiently.



\section*{Discussion}

We have demonstrated that deep reinforcement learning quantum compilers can approximate single-qubit gates by a set of quantum gates without prior knowledge. 

We point out that our method differs from existing quantum compiler algorithms for its irrespectiveness to the basis and its flexibility. Indeed, \textsc{Y-Z-Y} gate decomposition can only manage a basis consisting of y and z rotations, while \textsc{KAK} decomposition~\cite{vatan2004optimalKAK} is limited to two-qubits and \textsc{CNOT} and y and z rotations. Machine learning methods based on A*~\cite{hart1968formalAstar} algorithms could suffer from high execution time that could scale sub-optimally for high accuracy~\cite{davis2020towards, PhysRevLett.125.170501}.
Contrary to design a tailor-suited quantum compiling algorithm, we exploited a \textsc{DRL} agent to learn a general strategy to approximate single-qubit unitary matrices and store it within an artificial neural network.

One of the critical question to consider when measuring the performance of a quantum compiler is how a classical computer can efficiently return the sequence of gates. Inefficient strategies~\cite{lloyd:almostanyuniversal} can neutralize any quantum advantage over classical counterparts if the execution-time and the length of the sequence scale sub-optimally with the accuracy.

\begin{figure}
    \centering
    \includegraphics[width=0.6\textwidth]{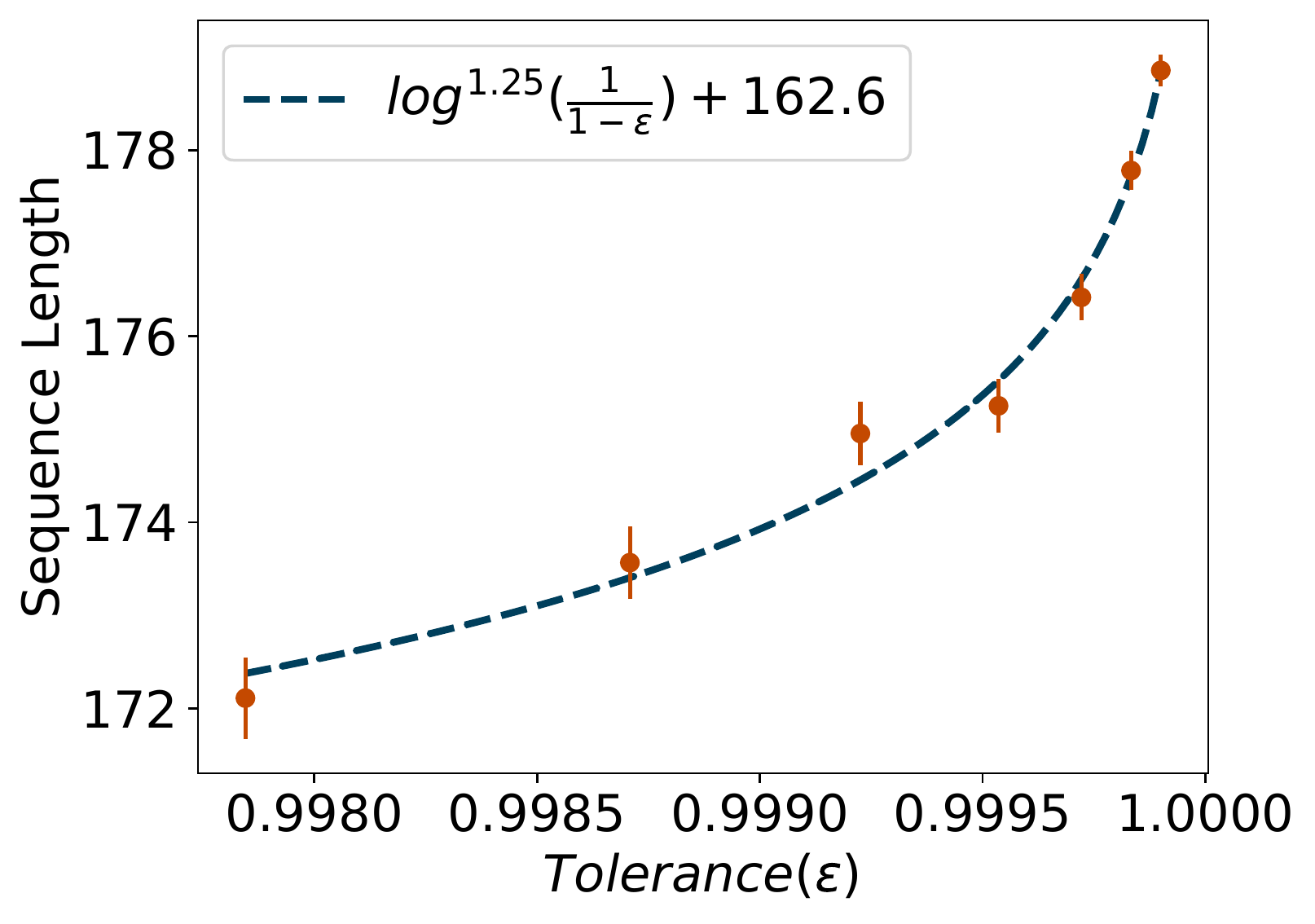}
    \caption{\textbf{Relation between the length and the tolerance}. Each data-point is obtained by averaging the length of the approximating sequence of gates found by a trained agent using a validation set of $10^7$ unitary targets. The agent was trained to achieve a final tolerance of $0.9999$ \textsc{AGF}. The targets are built as compositions of small rotations around the three axes of the Bloch sphere, as described in subsection~\num{3.1} of the Supplementary Note. The data fit a polylogarithmic function represented by the blue line almost perfectly as highlighted by the $R^2$ factor of $0.986$ and the \textsc{RMSE} of $0.26$.}
    \label{fig:performance}
\end{figure}

Our \textsc{DRL} quantum compiler can achieve sequences that scale as $\mathcal{O}(\log^{1.25}(1/\delta))$, as pointed by empirically measuring the performance on a specific task and shown in Figure~\ref{fig:performance}. Although such an approach has no guarantee to return a suitable solution, \textsc{DRL} quantum compilers can return solutions in minimal time, after a precompilation procedure to be performed once, hugely improving the execution time and enabling online quantum compilation. In fact, by writing the policy into a deep neural network, the execution time depends on the complexity of the network and the episode length only. Therefore, it scales proportionally to the sequence length, i.e. as $\mathcal{O}(\log^{1.25}(1/\delta))$. At the end of the training procedure, the agent returns the whole approximating sequence in a fraction of a second ($5.4 \cdot 10^{-4} s$ per time-step) on a single \textsc{CPU} core. The speed-up gain can be further enhanced by reducing the size of the neural networks and being easily parallelizable by exploiting specialized hardware to run them, such as \textsc{GPU}s or Tensor Processing Unit~\cite{dean:TPU}.

As noticeable examples, we trained and employed two deep reinforcement learning quantum compilers to build quantum circuits within a final tolerance of $0.99$ \textsc{AGF}, using two different sets of quantum logic gates. We accounted for the diverse characteristics of the bases designing a dense and a sparse reward functions. We addressed Haar distributed unitary matrices as targets to be as general as possible, but if additional information on the targets are available, it is possible to achieve higher tolerance without fine-tuning neural network architectures or the \textsc{RL} hyperparameters, as shown in the Supplementary Note~$3.1$.

Our method could be employed in larger qubit spaces, as shown by an early prototype in the Supplementary Note~$3.2$. The \textsc{DRL} compiler can approximate two-qubit logic gates with consistent performance compared to the one-qubit gates of $0.99$ \textsc{AGF}.

As a concluding remark, we observe that this approach can be specialized, taking into account any hardware constraints that limit operations, integrating them directly into the environments. These tests, as well as the extension of this approach to n-qubits, will be the objective of future work.

\section*{Methods}

\subsection*{Generation of Haar random unitary matrices}
The strategy used to generate the training data set should be chosen wisely, depending on the particular set of gates of interest, since deep neural networks are very susceptible both to the range and the distribution of the inputs. Therefore, Haar random unitaries have been used as training targets. Pictorially, picking a Haar unitary matrix from the space of unitaries can be thought as choosing a random number from a uniform distribution~\cite{russell:HUR}. More precisely, the probability of selecting a particular unitary matrix from some region in the space of all unitary matrices is directly proportional to the volume of the region itself. Such matrices form an unbiased data set which is ideal to train neural networks.

\subsection*{Learning by Hindsight Experience Replay (\textsc{HER)}}
Many \textsc{RL} problems may be efficiently addressed employing sparse rewards only, since engineering efficient and well-shaped reward functions can be extremely challenging. Such rewards are binary, i.e. the agent receives a constant signal until it achieves the goal. However, if the agent gets the same reward almost every time, it cannot learn any relationships of cause and effect that its actions have on the environment. Therefore, it might take an extremely long time to learn something, if anything at all. 

Hindsight Experience Replay (\textsc{HER}) is a technique introduced by \textsc{OpenAI} that allows to mitigate the sparse-reward problem~\cite{andrychowicz:HER}. The basic idea of \textsc{HER} is to exploit the ability that human have to learn form failure. Specifically, even if the agent always failed to solve the task, it can reach different objectives. Exploiting this information, it is possible to train the agent to reach different targets. Although the agent receives a reward signal to achieve a distinct goal from the original one, this procedure, if iterated, can help the agent to learn how to generalize the policy to reach the primary task we want to solve.

The implementation of \textsc{HER} in the Q-learning algorithm is straightforward. After an entire episode is completed, the experiences associated with that episode are modified selecting a new goal. Then, the q-function is updated as usual. There are several strategies to choose the goals as shown in ref.~\cite{andrychowicz:HER}. We designed a strategy to select the new goals, consisting in randomly selecting $k$-percent of the states which come from the same episode.

\subsection*{Average gate fidelity}
The average fidelity $\bar{F}(\mathcal{U}, U)$ between a two gates $\mathcal{U}$ and $U$ is defined by
\begin{equation}
    \bar{F}(\mathcal{U}, U) = \int \bra{\psi} \mathcal{U}^{\dagger} U \ket{\psi}\bra{\psi} U^{\dagger} \mathcal{U} \ket{\psi} d\psi, \qquad \int d\psi = 1
\end{equation}
where the integral is over all the state space using an Haar measure.

\subsection*{Neural network architectures}
The architecture of the deep neural network directly affects the performance of the \textsc{DRL} agent. However, choosing the optimal architecture is a trial and error task and can be an exceptionally time-consuming procedure, since it depends on the specific problem the agent is addressing. Therefore in this work, we didn't focus on the optimization, but on finding the smallest neural network architecture that can lead to satisfactory performance. We started with one hidden layer only and a few neurons, gradually increasing the depth and the width of the network. We found that a relatively small architecture made by two hidden layers of $128$ neurons is sufficient to achieve the tasks.

\subsection*{Software and Hardware}
All the code in this work was developed using Python language. The Stable Baseline~\cite{stable-baselines} library has been employed for the implementation of \textsc{PPO} agent only. Most of the simulation has been run by using \textsc{GNU} parallel~\cite{tange2011gnu} on an Intel Xeon W-2195 and a Nvidia GV100.

\section*{Data availability}
The data that support the findings of this study are available from the corresponding author upon reasonable request.

\section*{Code availability}
The code and the algorithm used in this study are available from the corresponding author upon reasonable request.

\section*{Acknowledgements}
L. M. and E. P. gratefully thank Vista Technology SRL for having partially supported this research. E.P. gratefully acknowledges the support of NVIDIA Corporation for the donation of the Titan Xp GPU used for this research.

\section*{Author contributions}
L.M. wrote all the codes and performed the experiments, M.P. contributed to the quantum mechanical environment of the RL agent, M.R. contributed to the development of the RL agents, E.P. conceived and coordinated this research. All the Authors contributed to discuss the results and to the writing of the manuscript. 

\bibliography{database}

\section*{Extended Figures and Tables}

\begin{table}
    \centering
    \begin{tabular}{cll}
        \toprule
        \textbf{Area related} & \textbf{Hyperparameter} & \textbf{Value} \\
        \midrule
        \multirow{3}*{Neural Network}       & \# hidden layers    & $128,\,128$ \\
                                            & activations         & SELU, SELU, linear \\
                                            & initializers        & lecun, lecun, glorot \\
        \midrule
        \multirow{4}*{Training}             & optimizer          & Adam    \\
                                            & learning rate      & \num{0.0005}  \\
                                            & batch size         & $10^3$      \\
                                            & training frequency & every $1$ episode \\
        \midrule
        \multirow{3}*{Algorithm}            & epsilon decay      & $0.99976$  \\
                                            & memory size        & $10^4$ experiences  \\
                                            & max length episode       & $130$ \\
        \bottomrule
    \end{tabular}
    \caption{List of the hyperparameters and their values used in the fixed-target problem.}
    \label{tab:fixedTarget}
\end{table}

\begin{table}
    \centering
    \begin{tabular}{cll}
        \toprule
        \textbf{Area related} & \textbf{Hyperparameter} & \textbf{Value} \\
        \midrule
        \multirow{2}*{Neural Network}       & \# hidden layers    & $128,\,128$ \\
                                            & activations         & SELU, SELU \\
        \midrule
        \multirow{3}*{Training}             & learning rate      & \num{0.0001 }  \\
                                            & batch size         & $128$      \\
                                            & \# agents           & 40 \\
        \midrule
        \multirow{1}*{Algorithm}            & max length episode       & $300$ \\
        \bottomrule
    \end{tabular}
    \caption{List of the hyperparameters and their values used by the PPO agent in the problem of quantum compiling by rotations operators.}
    \label{tab:rotationsHaar}
\end{table}

\begin{table}
    \centering
    \begin{tabular}{cll}
        \toprule
        \textbf{Area related} & \textbf{Hyperparameter} & \textbf{Value} \\
        \midrule
        \multirow{3}*{Neural Network}       & \# hidden layers    & $128,\,128$ \\
                                            & activations         & SELU, SELU, linear \\
                                            & initializers        & lecun, lecun, glorot \\
        \midrule
        \multirow{4}*{Training}             & optimizer          & Adam    \\
                                            & learning rate      & \num{0.0001}  \\
                                            & batch size         & $200$      \\
                                            & training frequency & every $1$ episode \\
        \midrule
        \multirow{3}*{Algorithm}            & epsilon decay      & $0.99931$  \\
                                            & memory size        & $5\cdot10^5$ experiences  \\
                                            & max length episode       & $130$ \\
        \bottomrule
    \end{tabular}
    \caption{List of the hyperparameters and their values used by a DQL agent in the \textsc{HRC} problem.}
    \label{tab:HRC}
\end{table}

\begin{table}
    \centering
    \begin{tabular}{ccccc}
        \toprule
        \textbf{Base} & \textbf{Solved (\%)} & \textbf{Mean Length} & \textbf{$95$\textsuperscript{th} percentile} & \textbf{$99$\textsuperscript{th} percentile} \\
        \midrule
        \textsc{HRC}                       & $95.0$  & $35$    & $94$  & $120$     \\
        \midrule
        Rotations                    & $96.4$  &  $124$  & $204$ & $245$     \\
        \bottomrule
    \end{tabular}
    \caption{Performance of the \textsc{PPO} and \textsc{DQL} agents after the training procedures in approximating Haar unitary matrices over a validation set of $10^6$ targets. The $95$\textsuperscript{th} and $95$\textsuperscript{th} columns refer to the $95$\textsuperscript{th} and $95$\textsuperscript{th} percentile of the distribution of the length of the solved sequences.}
    \label{tab:finalResults}
\end{table}

\end{document}